\documentclass[]{revtex4}
\usepackage{bm}
\usepackage{amssymb}
\usepackage{graphicx}
\usepackage{subfigure}
\newcommand{\Keywords}[1]{%
\begin{center}%
    Keywords: {#1} \\%
\end{center}%
}%

\begin{document}

\title{On the role of confinement on solidification in pure
materials and binary alloys}

\author{B. P. ATHREYA$^\dagger$, J. A. DANTZIG$^\dagger$\thanks{corresponding author, dantzig@uiuc.edu}, S.
LIU$^\ddagger$, and R. TRIVEDI$^\ddagger$}
\affiliation{$^\dagger$Department of Mechanical and Industrial Engineering,\\
University of Illinois, Urbana, IL 61801.\\
$^\ddagger$Department of Materials Science and Engineering, \\
Iowa State University, Ames, IA 50011.}


\begin{abstract}
We use a phase-field model to study the effect of confinement on
dendritic growth, in a pure material solidifying in an undercooled
melt, and in the directional solidification of a dilute binary
alloy. Specifically, we observe the effect of varying the vertical
domain extent ($\delta$) on tip selection, by quantifying the
dendrite tip velocity and curvature as a function of $\delta$, and
other process parameters. As $\delta$ decreases, we find that the
operating state of the dendrite tips becomes significantly
affected by the presence of finite boundaries. For particular
boundary conditions, we observe a switching of the growth state
from 3-D to 2-D at very small $\delta$, in both the pure material
and alloy. We demonstrate that results from the alloy model
compare favorably with those from an experimental study
investigating this effect.
\end{abstract}
\maketitle



\section{Introduction}

Dendrites are one of the basic microstructural patterns seen in
solidified metals. The mechanical behavior of the solidified
product is often decided by the length scales set by these
patterns. Study of dendritic growth is therefore motivated by the
need to predict these length scales. The fundamental quantities
that completely describe the growth of a dendrite at steady state
under a given set of external conditions are its tip velocity and
radius, which together define the so called ``operating state''.
Despite considerable advances in the understanding of
solidification science, discrepancies still arise when one
attempts to compare theoretical predictions of dendrite operating
states with experimental observations. We find that some of these
discrepancies derive from differences between the ideal conditions
assumed in theoretical treatments, and those experienced by
materials under actual experimental situations.

The first theoretical treatment of the ``free'' dendrite growth
problem was presented by Ivantsov \cite{Iva47}. He considered a
pure dendrite, modeled as a paraboloid of revolution with tip
curvature $\rho_{tip}$, growing into an infinite undercooled melt
with temperature $T \rightarrow T_\infty$ far from the advancing
tip. The dendrite was assumed to be isothermal at the melting
temperature $T_m$, and to be growing along its axis at constant
velocity $V_{tip}$ in a shape-preserving way. Ivantsov found a
solution to the thermal transport problem, in which  the
dimensionless undercooling $\Delta = (T_m-T_\infty)/(L_f/c_p) =
{\cal{I}}(Pe)$ where $Pe = \rho_{tip}V_{tip} / 2 D$ is the
P\'eclet number, $D$ is the thermal diffusivity, and ${\cal{I}}$
is the Ivantsov function. The temperature has been scaled by the
characteristic temperature $L_f/c_p$, with $L_f$ being the latent
heat of fusion and $c_p$ the specific heat.

This solution presented a conundrum, because it showed that the
transport problem alone did not uniquely specify the operating
state of the dendrite, i.e., the single combination of
$\rho_{tip}$ and $V_{tip}$ observed in experiments. Additional
considerations, such as the effect of curvature on the melting
point \cite{Tem60}, stability \cite{langer+mul1,langer+mul2} and
eventually the anisotropy of the surface tension
\cite{Bro83,Ben83,Ben84}, led to a second condition $ \sigma^* =
{2 d_0 D / \rho_{tip}^2 V_{tip}} $ where $\sigma^*$ is called the
\emph{selection constant}, and $d_0$ is the capillary length. The
combination of Ivantsov's solution (modified for surface tension
and its anisotropy) and the condition $\sigma^*$ is constant gives
a unique operating state. Numerical simulations using the
phase-field method \cite{Karma+Rappel} at large values of
$\Delta$, have found agreement with the predictions of this body
of work, known as \emph{microscopic solvability theory}.

Glicksman and co-workers developed experimental techniques for
studying the solidification of pure materials, with the objective
of observing the operating state. They performed experiments with
phosphorous \cite{glick+schaefer}, and transparent analog alloys
like succinonitrile (SCN) \cite{Hua81} and pivalic acid (PVA)
\cite{IDGE_des}. The results of these careful experiments found
some areas of agreement with microscopic solvability theory, in
particular, the value of $\sigma^*$ was found to be constant, but
the operating combination of $\rho_{tip}$ and $V_{tip}$ did not
agree. Provatas, \emph{et al.} were able to explain this
discrepancy by showing that for the low undercooling conditions
found in the experiments, interaction between neighboring dendrite
branches \cite{Pro98a,Pro98b} affected the operating state.

Experiments have also been performed to examine the role of
superimposed fluid flow on dendritic growth. Gill and coworkers
\cite{gill,Lee} used SCN in a special cylindrical chamber with a
bellows to effect fluid flow. Bouissou and Pelc\'{e}
\cite{bouissou_expt} performed experiments with a flowing alloy of
PVA and a seed confined between microscope slides. Saville and
Beaghton \cite{saville} presented a theoretical analysis which
extended Ivantsov's solution to consider the superimposed flow.
Jeong \emph{et al.} \cite{Jeong2001} performed phase-field
simulations of these experiments, and once again found
discrepancies with theory. They conjectured that the differences
arose because of the effect of finite containers in the
experiments, leading to boundary conditions which differed from
the assumptions of infinite media used in the theory.

Dendrite tip theories for constrained growth, such as directional
solidification of dilute binary alloys between microscope slides,
have been developed by Trivedi \cite{Trivedi80} and, Kurz and
Fisher \cite{Kurz+Fisher}. They have shown that a relationship of
the form $\rho_{tip}^2 V_{tip} = \mbox{constant}$, should hold for
constrained growth just as in free dendrite growth. Early
experiments by Somboonsuk \emph{et al.} \cite{Somboonsuk84} in
samples with slide separation greater than 150 $\mu\mbox{m}$ have
shown excellent agreement with this theory. However, in recent
studies Liu \emph{et al.} \cite{liu+trivedi} have demonstrated
that experimental results start to deviate significantly from
theory when the slide separation approaches the scale of the
primary dendrite spacing.

In this article, we systematically study the role of confinement
on dendritic growth. Since \emph{every} experiment is performed in
a finite container, we feel that this effect cannot be ignored.
The first two authors have previously reported a study on confined
growth in pure materials \cite{athreya+dantzig1}. Here we extend
our investigations to directionally solidified binary alloys. For
purposes of continuity and completeness, we will again present
results from our study on pure materials. For the binary alloy,
recent experimental data from Liu \emph{et al.}\cite{liu+trivedi},
will provide us with an avenue for testing our numerical
predictions.


\section{Modeling \label{s2}}

\subsection{Phase field model, adaptive grids and numerical methods}

The objective in a general solidification problem is to solve the
equations governing thermal and solute transport, subject to
boundary conditions on the solid-liquid interface (moving
boundary) and elsewhere. If melt convection is to be modeled, one
needs to solve the momentum equations for fluid flow
simultaneously with the above transport equations. Imposing the
interface boundary conditions upon discretizing the governing
equations poses a difficulty however, since the interface, as it
evolves, will not in general align itself with a fixed set of mesh
points.

The phase-field method eliminates the sharp liquid-solid boundary
by introducing evolution equations for a continuous order
parameter $\phi \in [-1,1]$, where $\phi=-1,+1,0$ corresponds to
liquid, solid and interface respectively. Thus, the arduous task
of solving the transport equations separately in liquid and solid
domains while simultaneously satisfying boundary conditions on
arbitrarily shaped interfaces, is replaced by that of solving a
system of coupled differential equations; one for the evolution of
$\phi$ and one for each of the transport variables (temperature,
concentration and velocity). Phase-field modeling has been an
active area of research in the past decade, and we refer the
interested reader to original work by Langer \cite{Langer86},
Karma and Rappel \cite{Karma+Rappel}, and Beckermann \emph{et al.}
\cite{Beckermann1} for derivations of the phase-field equations
and selection of phase-field parameters ensuring convergence to
the original sharp interface problem.

The phase-field model introduces a parameter $W_0$ that connotes
the finite width of the now `smeared' interface. Karma and Rappel
\cite{Karma+Rappel} showed that the model converges to the sharp
interface equations when $p = W_0V_{tip}/D \ll 1$, where $D$ is
the thermal or concentration diffusion coefficient, and $V_{tip}$
is the nominal tip velocity of the dendrite. Resolving the
interface on a discrete mesh requires that the mesh spacing
$\Delta x \sim W_0$, while demanding that the diffusion field not
interact with the boundaries leads to the domain size $L_B \gg
D/V_{tip}$. Satisfying these requirements causes calculations on
regular meshes to quickly reach the limit of available computing
resources. For example, if we choose $p = 0.01$, fix $L_BV_{tip}/D
= 10$, and enforce $\Delta x = W_0$, then we find that the number
of grid points per dimension on a regular mesh should be at least
$L_B/W_0 = 1000$. This makes computations challenging on regular
meshes even in 2-D, while 3-D computations may not be practical at
all, depending on available computing power.

We have mitigated this problem successfully by solving the
equations on an adaptive finite element mesh
\cite{Pro98b,Jeong2001}. In three dimensions, we use eight-noded
trilinear brick elements stored using an octree data structure. A
local error estimator indicates refinement or coarsening of the
mesh, and this permits tracking of the interface as well as
resolution of gradients in the other fields. There are six degrees
of freedom at each node (three velocities, pressure,
temperature/concentration and $\phi$), and a typical computation
reaches well over one million unknowns. The finest elements
($\Delta x_{min}$), which are distributed near the interface, now
need to be order of $W_0$.

For our studies on pure materials, we have used a finite element
discretization of the 3-D phase field model developed by Karma and
Rappel \cite{Karma+Rappel}. In order to account for the effects of
melt convection we adopt the formulation presented by Beckermann
\emph{et al.} \cite{Beckermann1}, who use an averaging method for
the flow equations coupled to the phase-field. By appropriate
choice of phase-field parameters we have ensured zero interface
kinetics, which is a valid assumption for the range of
undercooling we are concerned with.

For our alloy simulations, we have used a one sided
(vanishing solid diffusivity) phase-field model
\cite{Karma2001,Ramirez04,Echebarria04}, with a frozen temperature
approximation. In a directional solidification arrangement, for
certain values of the problem parameters (particularly when
simulating real materials), a considerable amount of time can
elapse before the transients vanish and the solid-liquid interface
reaches steady state. In that time, the interface can encounter
the end of the simulation box if the equations are solved in a
reference frame that is fixed globally, and if the box is not
large enough to contain the diffusion field. To alleviate this
difficulty, we have solved the phase-field equations in a
coordinate frame translating with the pulling speed. This saves
some computational expense by allowing us the use of smaller
boxes. We have not investigated the effect of melt convection in
our numerical experiments with alloys.

In a recent article, Echebarria \emph{et al.} \cite{Echebarria04}
have emphasized that the same choice of phase-field parameters
that produced zero interface kinetics in the pure material cannot
also ensure this condition in the alloy model. This is due to the
presence of certain additional terms in the kinetic parameter
$\beta$, that arise out of accounting for the discontinuity in the
concentration field at the interface in a model with vanishing
solid diffusivity. To ensure that the kinetic coefficient is
negligible at the interface, the phase-field relaxation time
$\tau$ needs to be made temperature dependent in this region by
setting
\begin{equation}
\tau = \tau_0\left[1-(1-k)\frac{z - V_pt}{l_T}\right].
\end{equation}
Here $\tau_0$ is the usual relaxation time, $k$ is the partition
coefficient, $z$ is the distance from the interface, $V_p$ is the
pulling velocity, $t$ is time and $l_T$ is the thermal length
\cite{Karma2001}.

We evolve the nonlinear order parameter equation using a
Forward-Euler time stepping scheme, while the linear
thermal/solute transport equations are solved using the
Crank-Nicholson scheme with a diagonally preconditioned conjugate
gradient solver. The transport equations typically converge in
fewer than five iterations per time step. The 3-D flow equations
for the pure material are solved using the semi-implicit
approximate projection method (SIAPM) \cite{Gresho1995}. Details
of the above numerical methods and the finite element formulation
are omitted here, as they have been presented elsewhere
\cite{Jeong2001}.

\subsection{Geometry, initial and boundary conditions}

\begin{figure}[htbp]
\begin{center}
{\includegraphics[height=4.0in,angle=0]{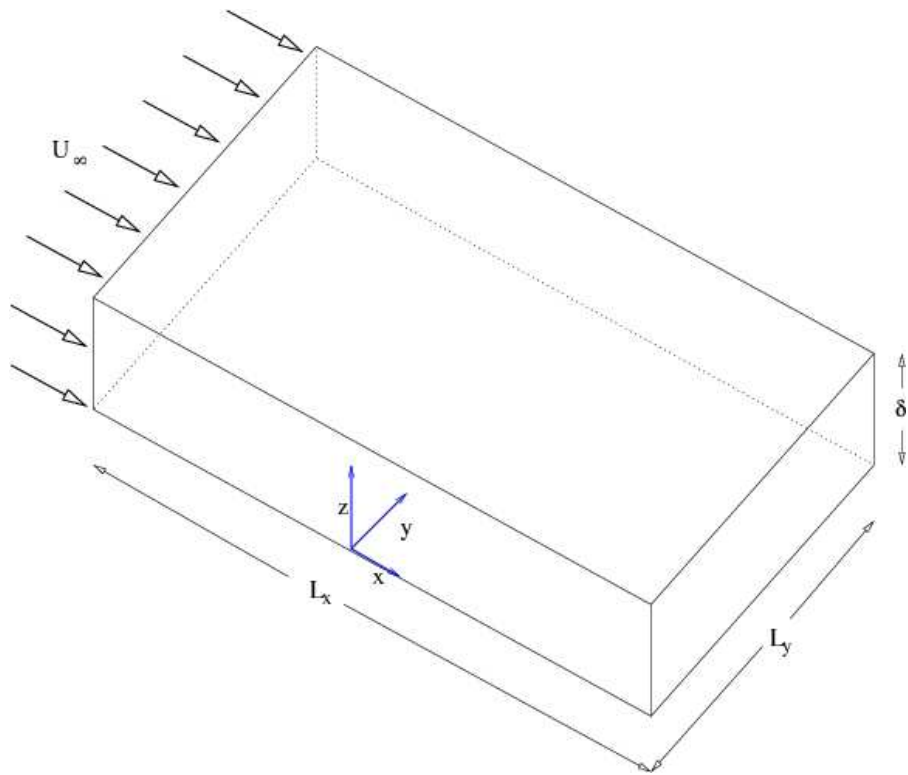}}
\caption{\label{box} Simulation Domain. All surfaces are modeled
as symmetry planes in pure material simulations, whereas the
surfaces $y=0$ and $y=L_y$ are periodic boundaries in the alloy
simulations.}
\end{center}
\end{figure}

Our three dimensional simulation domain is the rectangular
parallelepiped illustrated in Fig. \ref{box}, with edge lengths
along the \emph{x}, \emph{y} and \emph{z} axes; $L_x$, $L_y$ and
$\delta$ respectively. The edges of the box are oriented along
$\langle100\rangle$ cubic crystallographic directions.


For the pure material, the initial condition is a spherical solid
seed with a radius greater than the critical nucleation radius,
centered at the origin depicted in Fig. \ref{box}. Because of the
inherent symmetry in the growth of the seed, it is usually
sufficient to model one octant in three dimensional space.
However, if forced fluid convection is incorporated along a
particular direction, then the solidification rates into and
counter to this direction become unequal. For example, if there is
a flow parallel to the \emph{x} axis, $x=0$ is no longer a plane
of symmetry. To account for this break in symmetry, we have to
model at least a quadrant of space, with $z = 0$ and $y = 0$ as
planes of symmetry.

In the simulations with the pure material, the dimensionless thermal
field $u$ is subjected to zero flux ($\nabla u \cdot \mathbf{n} = 0$)
boundary conditions on all surfaces. Fluid flow, when it is included
in our study, is imposed as an inlet boundary condition $U_{\infty}$,
normal to the face $x = -L_{x}/2$. The velocity field is subjected to
symmetry boundary conditions on the domain walls, and is forced to
vanish in the solid ($\phi = 1$) by an appropriate formulation of the
momentum equations (see \cite{Beckermann1}). We fix the lateral
dimensions of the simulation box ($L_x = 512$ and $L_y = 256$ are
typical values), and study the interface evolution as a function of
$\delta$, which is varied from 128 to 4. Here, $L_x$, $L_y$ and
$\delta$ are in units of the interface width $W_0$.  For very small
$\delta$ ($\le 8$), steady growth conditions are reached relatively
quickly for large undercooling. To save on computational cost in
these runs (where $\Delta x_{min} = 0.5$), we sometimes use shorter
lengths for $L_x$ and $L_y$, chosen to ensure that the diffusion
field does not interact with the ends (in the \emph{x} and \emph{y}
directions) of the box. For smaller $\Delta$ however, it typically
takes much longer to reach steady conditions, and when melt
convection is included, it can take impractically long CPU times to
get converged results. For these cases, we terminate our runs when
the tip radius/velocity versus time curves start to even out.
Fortunately, it turns out that the behavior we are interested in
appears for combinations of $\delta$ and $\Delta$ where steady state
conditions are always achieved.

In the alloy simulations, the initial condition is a planar
interface at $x = X_0$, perturbed by randomly spaced finite
amplitude fluctuations. The box in these simulations is taken to
represent the shallow channel between microscope slides where
directional solidification conditions are imposed, viz. a fixed
thermal gradient moving at a constant speed $V_p$. Once again, in
this arrangement we study the influence of the depth of the
channel $\delta$ (or equivalently the sample film thickness), on
interface morphology. To minimize the diffusion field's
interaction with the lateral boundaries, $L_x$ and $L_y$ are
chosen to be relatively large ($\sim 256$). We enforce zero flux
boundary conditions on the concentration field, on the surfaces
$x=L_x/2=-L_x/2$ and $z=0=\delta$, while periodic boundary
conditions are imposed on the boundaries $y=0$ and $y=L_y$. The
rationale behind periodic boundary conditions is to be able to
simulate an infinite domain in \emph{y}.

Unless otherwise stated, on each boundary, we employ the same type
of boundary condition on the phase-field variable $\phi$, as we do
on the transport variable. Where $\nabla \phi \cdot \mathbf{n} =
0$, the material ``wets'' the boundaries, and the corresponding
contact angle is $90^{o}$. S\'{e}moroz \emph{et al.} have
previously used this technique to capture wetting of solid
surfaces, with a two-dimensional phase-field model for binary
alloys \cite{rappaz2000}. We also show a calculation with $\phi =
-1$ on the boundaries, which is equivalent to making the material
``non-wetting'' (contact angle = $0^{o}$). The real contact
condition probably lies somewhere in between these two extremes.

\section{Effect of small $\delta$ in a pure material \label{s3}}

In this section, we report the effect of changing $\delta$ on the
tip of a pure material dendrite, evolving along the negative
\emph{x} axis (upstream direction when flow is present). We
simulated the cases shown in Table 1. We used a fixed value for
the four-fold anisotropy in all our simulations ($\epsilon_{4} =
0.05$). We have not corrected for grid anisotropy
\cite{Karma+Rappel} in these calculations, but work at a grid
spacing where its effect is known to be small \cite{Pro98b}.


\begin{table}{\centering Table 1. Pure material simulations}\\[0.5ex]
\setlength{\tabcolsep}{12mm}
\begin{center}
\begin{tabular}{llll} \hline\hline
Case & $\Delta$ & $U_{\infty}$ & $\delta$\\[0.1ex]
\hline
1(a)-(f) & 0.55 & 0         & 128,64,32,16,8,4\\
2(a)-(f) & 0.55 & 5         & 128,64,32,16,8,4\\
3(a)-(e) & 0.25 & 0         & 128,64,32,16,8\\
4(a)-(f) & 0.25 & 5         & 128,64,32,16,8,4\\
5(a)-(d) & 0.15 & 0         & 128,64,32,16\\
6(a)-(e) & 0.15 & 5         & 128,64,32,16,8\\[0.2ex]
\hline
\end{tabular}
\end{center}
\end{table}

\noindent We use the following values for parameters in our calculations: interface
width $W_0 = 1$, time scale for interface kinetics $\tau_0 = 1$, coupling constant
$\lambda = 6.383$, thermal diffusivity $D = 4$, capillary
length $d_{0} = 0.1385W_0$ (which leads to zero interface kinetics), and Prandtl number
$ Pr= 23.1$, where $W_0$, $\tau_0$ and $D$ are in dimensionless units (see
\cite{Karma+Rappel}). 

\subsection{``Wetting'' boundary conditions, $\nabla \phi \cdot \mathbf{n} = 0$}

\begin{figure}[htbp]
\begin{center}
{\includegraphics[height=4.0in,angle=0]{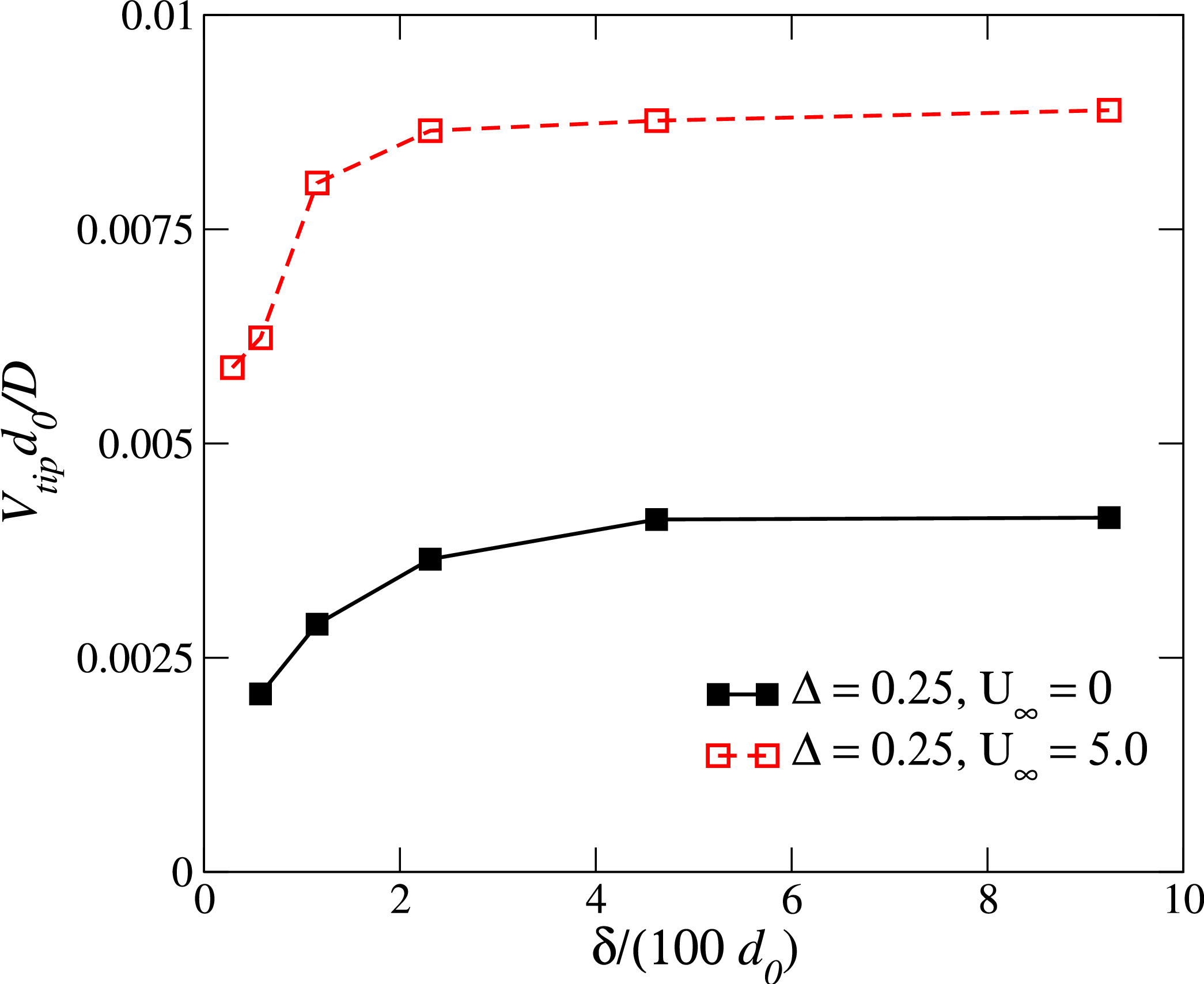}}
\caption{\label{vel_delta} Tip velocity vs. $\delta$,
corresponding to cases 3 and 4 in Table 1.}
\end{center}
\end{figure}

\begin{figure}[htbp]
\begin{center}
{\includegraphics[height=4.0in,angle=0]{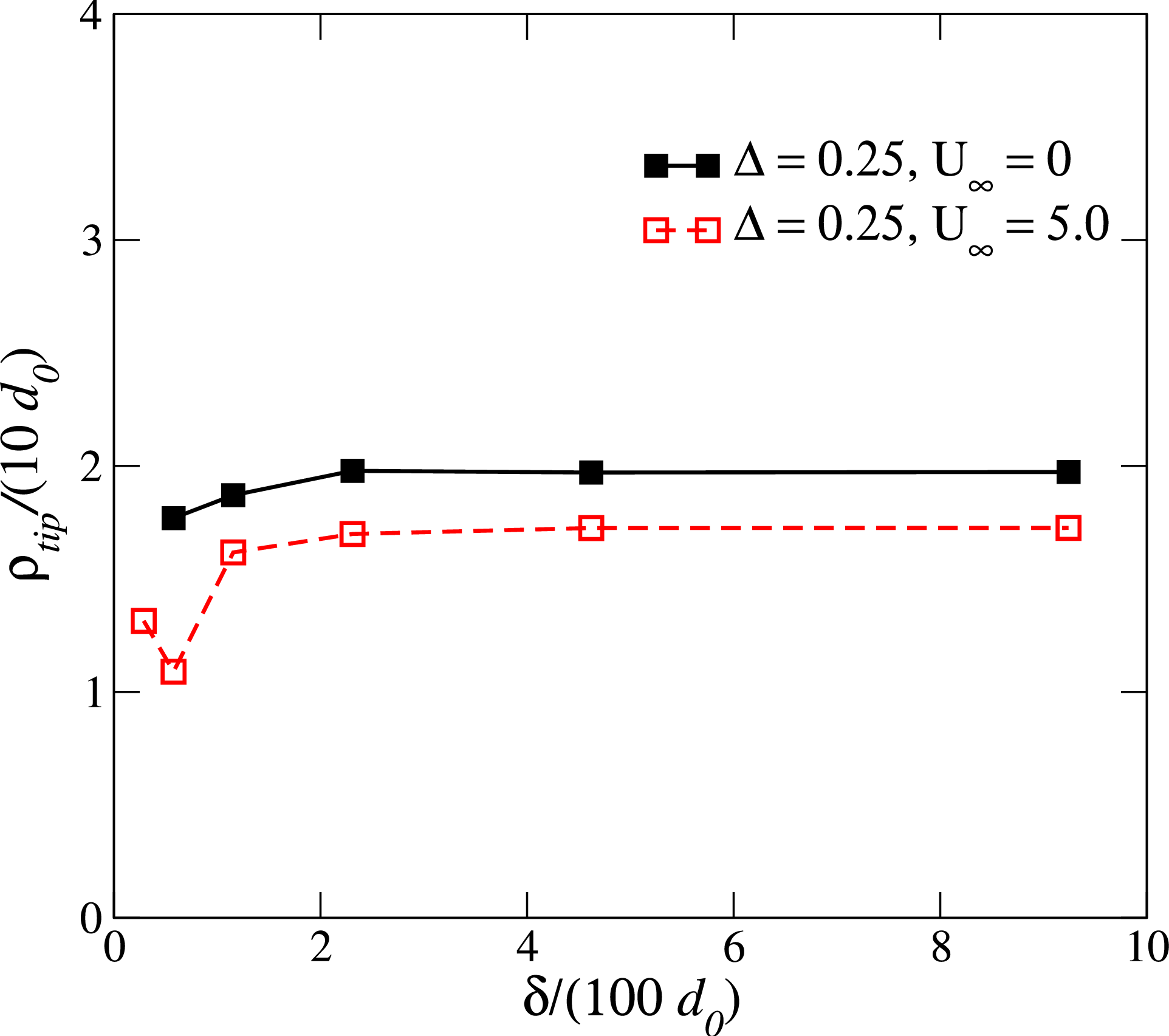}}
\caption{\label{rad_delta} Tip radius vs. $\delta$, corresponding
to cases 3 and 4 in Table 1.}
\end{center}
\end{figure}

To make ideas more concrete, we choose cases 3 and 4 as a
representative subset of our computations and present detailed
analyses on those runs. Figures \ref{vel_delta} and
\ref{rad_delta} show the upstream dendrite's tip velocity and
radius respectively, as functions of the box height $\delta$.


In these plots, $\delta$ and $\rho _{tip}$ were made dimensionless
by scaling them with $d_{0}$, and $V_{tip}$ is scaled by
$D/d_{0}$. We compute the tip radii $\rho _{xz}$ and $\rho _{xy}$
along two principal planes using the method of Jeong \emph{et al.}
\cite{Jeong2001}, and estimate the mean tip radius by the formula
\begin{equation}
\label{radcurv}\rho_{tip} = 2\left(\frac{1}{\rho_{xy}} +
\frac{1}{\rho_{xz}}\right)^{-1}.
\end{equation}
We can see from Fig. \ref{vel_delta} that for large values of
$\delta$, the tip velocity remains relatively unaffected by the
box height. However, as we go to  very small heights $V_{tip}$
decreases quite dramatically. As $\delta$ is decreased, there is
also a gradual decrease in the tip radius $\rho _{tip}$. Clearly
enough, box height has a pronounced effect on tip dynamics. Fluid
flow induces a parallel shift in these curves. The dendrite tip
velocity increases uniformly in the presence of flow
\cite{Jeong2001}. On the other hand, the tip radius is lower than
the case with pure diffusion.


\begin{figure}[htbp]
\begin{center}
\subfigure[\label{cplot_64}$\delta = 64$ corresponding to Case
4(b). \emph{u} ranges from A $\equiv$ -0.24 to L $\equiv$ 0.008]
{\includegraphics[height=1.6in,angle=0]{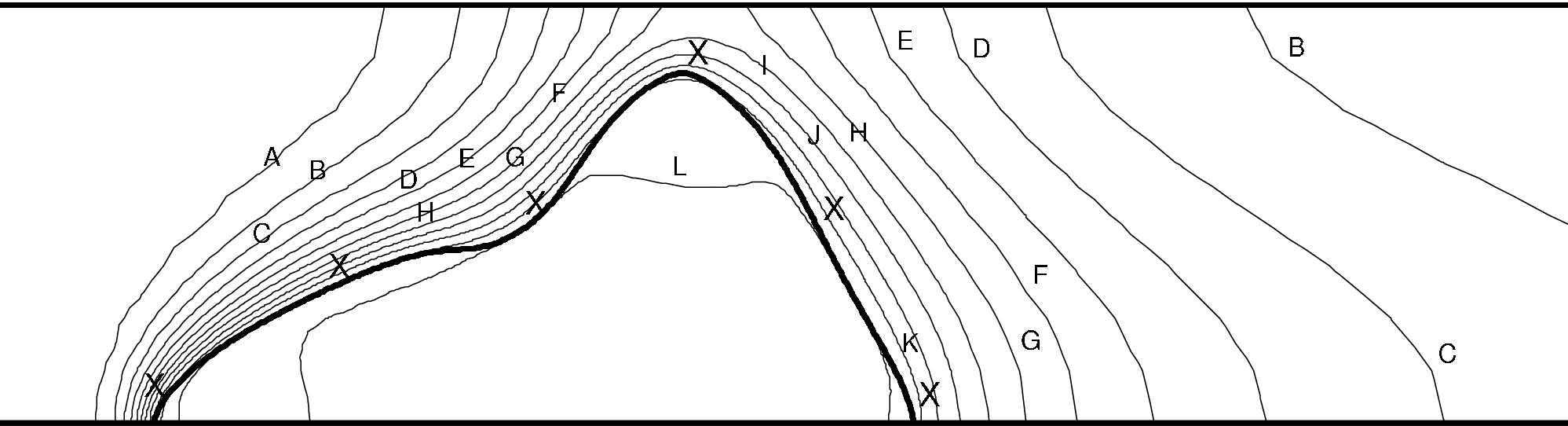}}
\subfigure[\label{cplot_8}$\delta = 8$, corresponding to Case
4(e). \emph{u} ranges from A $\equiv$ -0.24 to L $\equiv$ 0.008]
{\includegraphics[height=0.63in,angle=0]{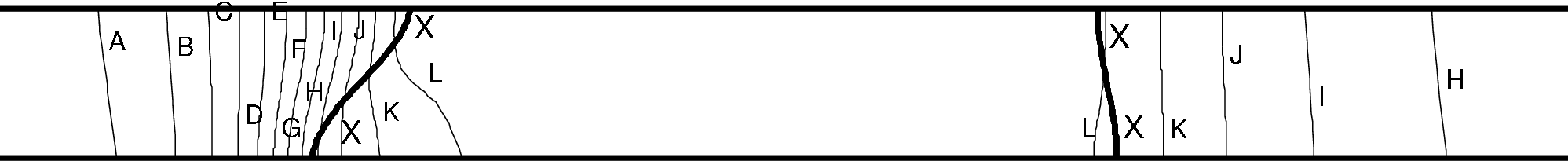}}
\caption{\label{cplot}Temperature contours for two different box
heights when $\Delta = 0.25$ and $U_{\infty} = 5.0$. In each case,
the letter X symbolizes the dendrite outline (bold contour), under
steady growth conditions. Contours are plotted in intervals of 0.025.
Notice that the contours near the dendrite tip are more spread out
when $\delta$ is smaller.}
\end{center}
\end{figure}

The observed trends can be explained as follows. As long as
$\delta$ is sufficiently large, the thermal field enveloping the
dendrite will interact with the upper boundary at a distance that
is relatively far behind the tip. In particular, the thickness of
the thermal boundary layer near the tip remains unaffected by this
interaction. However, as $\delta$ is decreased, this thickness can
grow quite rapidly. We illustrate this effect by examining the
temperature profile in the \emph{x-z} plane, as shown in Fig.
\ref{cplot}. It is evident that the temperature contours are more
spread out in Fig. \ref{cplot_8} where $\delta = 8$, compared to
those in \ref{cplot_64}, where $\delta = 64$. The increased
boundary layer thickness, decreases the thermal gradient into the
liquid at the liquid-solid interface, which in turn retards the
growth rate as a direct consequence of the Stefan condition. Due
to the zero flux boundary condition on the plane $z=\delta$,
further reduction in $\delta$ makes heat transfer in the vertical
direction almost completely ineffective. Tip curvature in the
$x-z$ plane vanishes and the dendrite switches morphology from 3-D
to 2-D. We note that once the dendrite goes 2-D, $\rho_{tip} =
\rho_{xy}$.


\begin{figure}[htbp]
\begin{center}
\subfigure[$\delta = 16$, 3-D dendrite]
{\includegraphics[height=2.1in,angle=0]{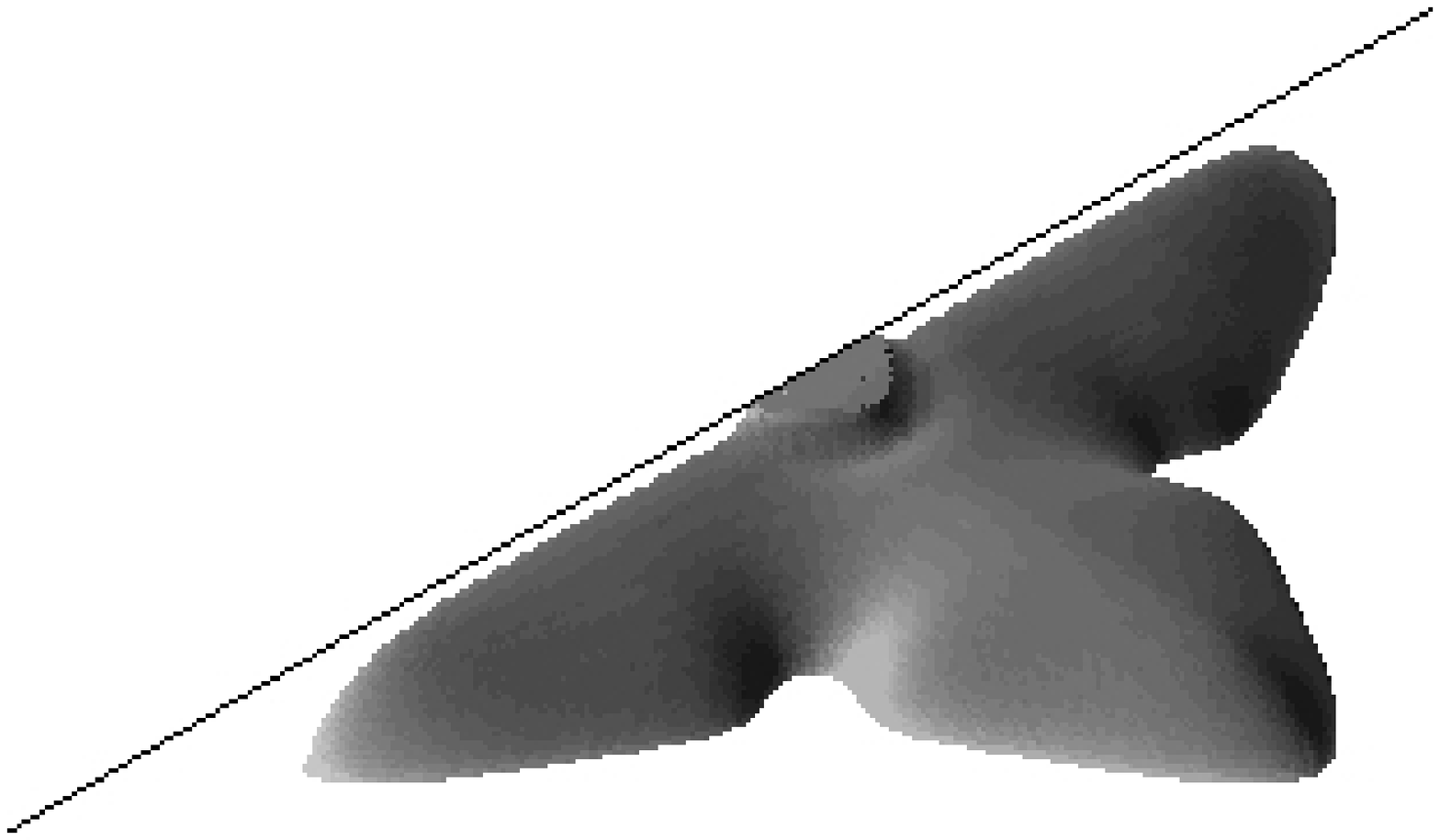}}
\hfill \subfigure[$\delta = 4$, 2-D dendrite]
{\includegraphics[height=2.0in,angle=0]{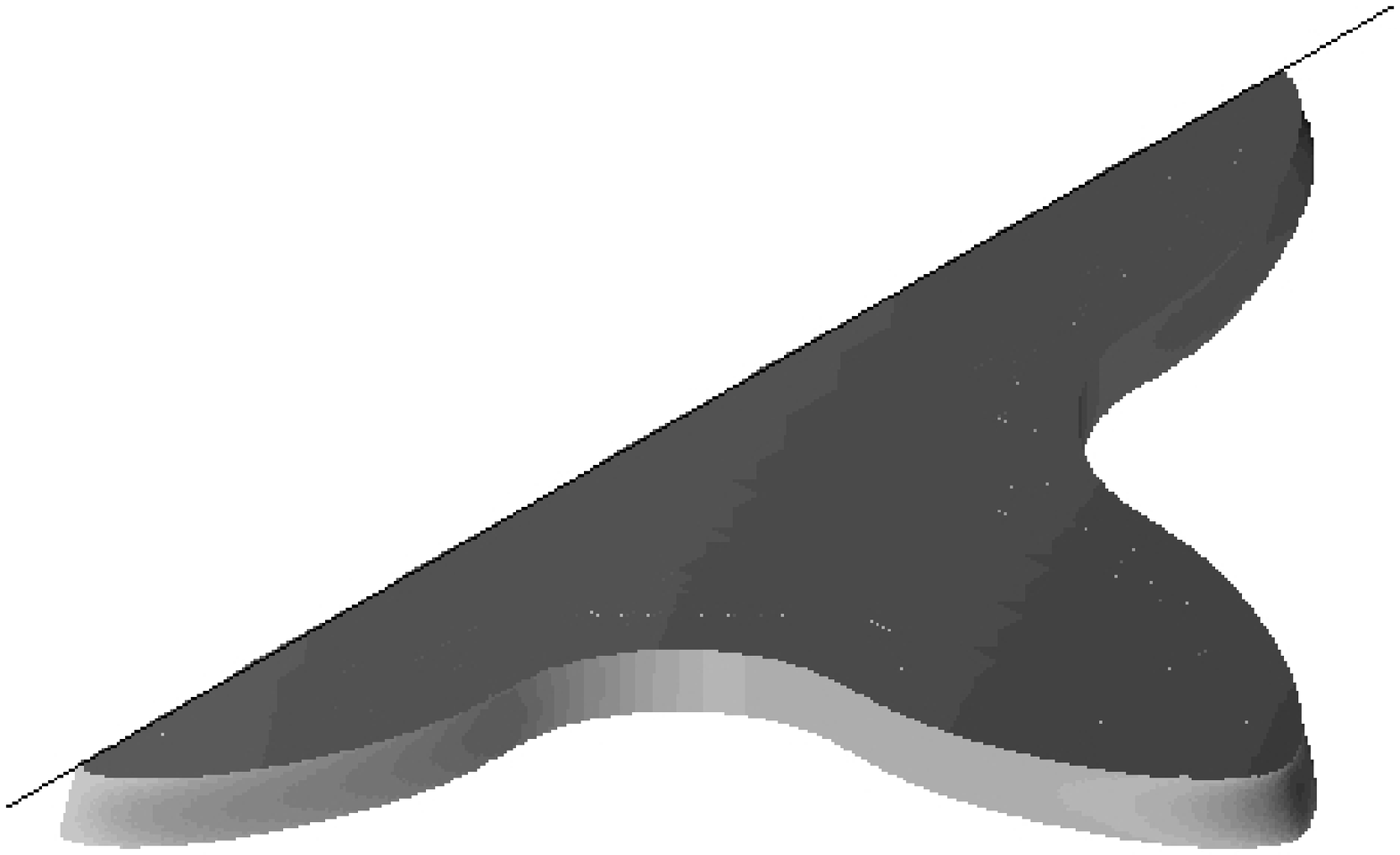}}
\caption{\label{3d_2d} 3-D to 2-D dendritic transition at small
$\delta$. The shaded surface is the dendrite ($\phi =0$). For
these runs $\Delta = 0.55$ and $U_{\infty} = 0$.}
\end{center}
\end{figure}

An interesting result here is the 3-D to 2-D transition. We have
performed tests with finer meshes (more elements in the vertical
direction) to ensure that it is not simply an artifact of poor
grid resolution. We believe this phenomenon to be a consequence of
the $\nabla \phi \cdot \mathbf{n} = 0$ boundary condition on the
upper boundary. The only way for the solid-liquid interface to
match this boundary condition at small $\delta$ is for the
curvature in the \emph{x-z} plane to vanish. This is illustrated
in Fig. \ref{3d_2d}.

\begin{figure}[htbp]
\begin{center}
{\includegraphics[height=4.0in,angle=0]{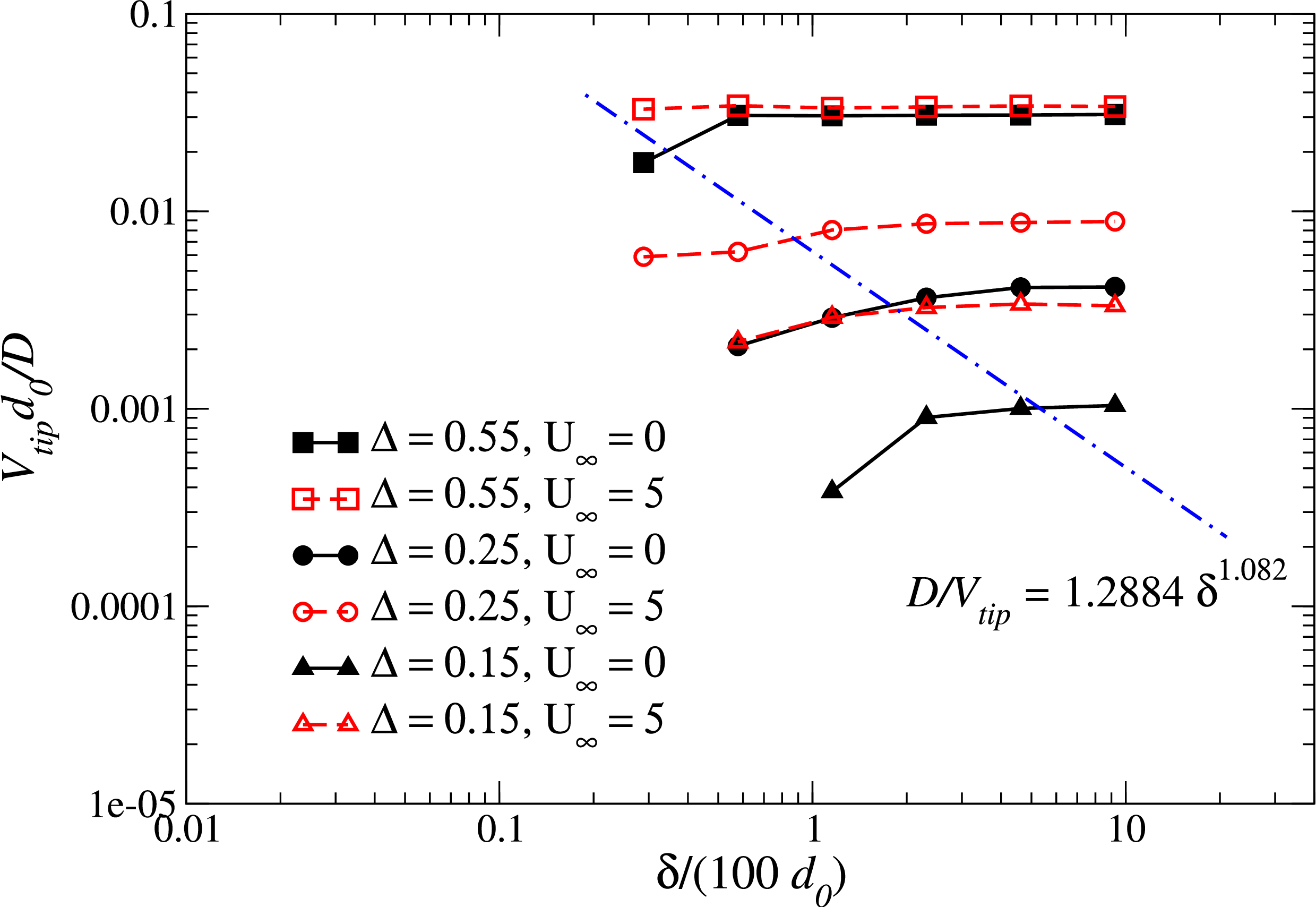}}
\caption{\label{vtip_delta} Tip velocity as a function of $\delta$
for different undercooling and flow conditions. A weak power law
relationship emerges between the $\delta$ below which interaction
effects are strong, and diffusion length $D/V_{tip}$.}
\end{center}
\end{figure}

It is also interesting to observe the effect that melt flow has on
the interaction between the tip and the boundary. We find that
melt convection reduces the value of $\delta$ where the tip
velocity deviates from its nominal value. A straightforward
explanation for this effect is that advection increases the rate
of heat transport from the upstream dendrite arm. This increases
the growth rate (hence $V_{tip}$), while compressing the boundary
layer, whose thickness scales as $D/V_{tip}$. Thus the dendrite
remains three dimensional for a smaller value of $\delta$ than was
previously possible, with convection absent. A more negative value
of the undercooling also has qualitatively the same effect on the
strength of the tip-boundary interaction, since $V_{tip}$ again
increases in this case. Fig. \ref{vtip_delta} summarizes these
observations succinctly. Both melt convection and larger
undercooling, cause points on respective curves whereupon
interaction effects become important, to shift to the left.


Using a graph such as Fig. \ref{vtip_delta}, one can derive a
semi-quantitative estimate as to when the operating state becomes
affected by the finite height of the container. If one assumes
that the tip velocity at $\delta = 128$ for each case is
approximately the tip velocity of a dendrite growing under
identical conditions in an infinite domain, it is possible to
quantify the influence on the operating state in terms of a
percentage deviation in the true tip velocity from this nominal
value. If, for example, we consider deviations of the order of 3\%
to constitute a change in the operating state, a least squares fit
to these cut-off points on the respective curves in Fig.
\ref{vtip_delta} yields the criterion
\begin{equation}
\label{criterion} \delta \ge
0.7912\;\left(\frac{D}{V_{tip}}\right)^{0.9242}.
\end{equation}
If this condition is not satisfied, then it is likely that the
operating state is being influenced by the boundary. A simple
condition such as the one in Eqn. (\ref{criterion}) may be used as
a rule of thumb in determining if experimental studies on free
dendrite growth, in geometries similar to ours, are free from
contamination. In fact, one may have intuitively guessed a
condition of the type $\delta \ge \alpha\;(D/V_{tip})$ (where
$\alpha$ is some constant) to apply based on physical arguments
alone, and Eqn. (\ref{criterion}) supports this conjecture.

\subsection{Non-wetting boundary conditions}

\begin{figure}[htbp]
\begin{center}
{\includegraphics[height=2.0in,angle=0]{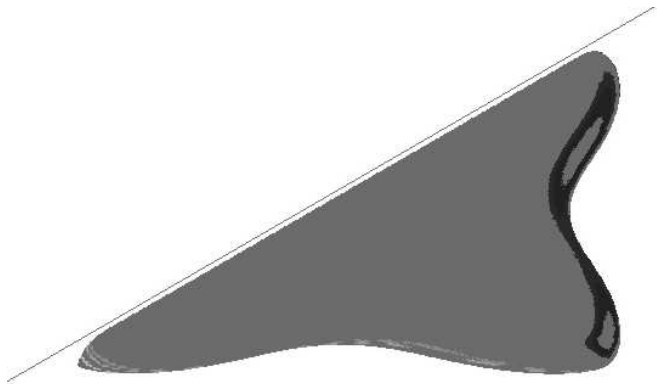}}
\caption{\label{dirichlet} Three dimensional dendritic growth for
$\Delta = 0.55$, $U_{\infty} = 0$  and $\delta = 4$, with
$\phi=-1$ on the upper boundary. Note that the dendrite does not
wet this surface.}
\end{center}
\end{figure}

To underscore the importance of phase field boundary conditions in
selecting a particular growth state, we present results from
another simulation with $\Delta = 0.55$, $U_{\infty} = 0$  and
$\delta = 4$. This time we impose $\phi = -1$ on the upper
boundary, which  corresponds to a physical situation where the
solidifying material is not allowed to wet the surface. A three
dimensional surface plot of a steadily growing dendrite is shown
for this case in Fig. \ref{dirichlet}.


As expected, the dendrite does not adhere to the top surface at
all. We conclude that this will be the case for any $\delta$ if
Dirichlet conditions of this nature are imposed. It is evident
then, that the 3D-2D transition that we saw previously is a strong
function of the boundary conditions imposed on the phase field. It
would be interesting to conduct a detailed investigation of the
effect of different boundary conditions on the tip in 3-D. The
interested reader is referred to the article by S\'{e}moroz
\emph{et al.} \cite{rappaz2000}, which discusses the influence of
different contact angles on a dendrite's  tip velocity, in
two-dimensional thin film solidification.

\section{Effect of small $\delta$ in a directionally solidified alloy \label{s4}}

In this section, we describe results from our simulations on a
directionally solidified alloy, and make comparisons with the
experimental data of Liu \emph{et al.} \cite{liu+trivedi}. These
simulations were conducted for the range of pulling speeds and
channel depths shown in Table 2. The depth $\delta$ was varied
from 64 to 4, and in each case, the simulation was continued until
the interface became stationary in the moving frame. From the
stable array of cells formed, a few were chosen as representative
of the array, their tip radii were extracted in the two principal
planes using a least squares quadratic polynomial fit, and the
mean tip radius of each cell was computed using Eqn.
(\ref{radcurv}). An average over these values was taken to be the
mean tip radius of the interface. We note that this process
required some approximation, especially in cases where we found
steadily growing cells of disparate sizes that would have made
such averaging inappropriate. In such cases, our usual approach
was to choose cells farthest from the boundaries, and where even
this failed to provide clear-cut choices, smaller cells were
chosen because of their lesser likelihood to split. In a majority
of our runs however, the choices were unambiguous. For certain
values of $\delta$ we found interface evolution to occur in a way
that cells would creep on either the top or bottom surfaces. In
those cases, the phase-field boundary conditions on the respective
surfaces allowed us to treat them as symmetry planes for the purpose of
calculating tip curvature.


\begin{table}{\centering Table 2. Alloy simulations}\\[0.5ex]
\setlength{\tabcolsep}{12mm}
\begin{center}
\begin{tabular}{lll} \hline\hline
Case & $V_p$ & $\delta$\\[0.1ex]
\hline
1 & 0.8 & 64,32,16,8,4\\
2 & 1.0 & 64,32,16,8,4\\
3 & 1.5 & 64,32,16,8,4\\
4 & 2.0 & 64,32,16,8,4\\[0.2ex]
\hline
\end{tabular}
\end{center}
\end{table}

\subsection{Selection of simulation parameters}

The following values for the lateral dimensions were seen to yield
satisfactory results. $L_x = L_y = 256$ when $\delta \le 16$, and
$L_x = 256$, $L_y = 128$, otherwise. We chose our simulation
parameters to keep computations tractable. It took about 90 hours
of CPU time on a 3.1 GHz processor to simulate a typical
directional solidification experiment for a chosen set of
phase-field parameters on a mesh with about 170 000 elements
($\delta = 64$). The interface required about 250 dimensionless time units
to reach
steady state in this case. As noted earlier, the use of a moving
reference frame allowed us to cut substantial costs associated
with the need for larger domains to prevent the diffusion field
from running out of the domain.

We did not attempt to model a real material in this study as this
caused our simulations to become considerably more expensive. To
illustrate this, consider a SCN-Salol system having the properties
listed in Table 3. The conditions in a directional solidification
experiment are completely described by the following two
dimensionless control parameters, $M = d_0/l_T = 6.66 \times
10^{-5}$ and $S = V_pd_0/D = 2.04 \times 10^{-4}$, where $d_0$ is
the capillary length, $l_T$ is the thermal length and $D$ is the
solute diffusivity in the liquid phase; for a pulling velocity of
$V_p = 5 \mu$m/s. To get converged results with the phase-field
model we require that the solution become independent of the
parameter $\epsilon = W_0/d_0$. After ensuring vanishing interface
kinetics, the following relationships involving the phase-field
parameters are realized: $D\tau_0/W_0^2 = a_1a_2\epsilon$,
$V_p\tau_0/W_0 = Sa_1a_2\epsilon^2$, and $l_T/W_0 = 1/(\epsilon
M)$. Here, $a_1 = 0.8839$ and $a_2 = 0.6267$, are constants that
arise in the phase-field formulation. \cite{Echebarria04} 


\begin{table}{\centering Table 3. Physical properties of a SCN-Salol alloy system}\\[0.5ex]
\label{scn_props}
 \setlength{\tabcolsep}{12mm}
\begin{center}
\begin{tabular}{ll}
\hline
$|m|$ (Liquidus slope) & 0.7 K/wt. \% \\
$D$ (Diffusion coefficient) & $8 \times 10^{-10}$ $\mbox{m}^2$/s \\
$\Gamma$ (Gibbs-Thomson coefficient) & $0.64 \times 10^{-7}$ K m \\
$k$ (Partition coefficient) & 0.2\\
$G$ (Thermal gradient) & 4 K/mm \\
$d_0$ (Capillary length) & $3.265 \times 10^{-8}$ m\\
$l_T$ (Thermal length) & $4.9 \times 10^{-4}$ m\\[0.2ex]
\hline
\end{tabular}
\end{center}
\end{table}

Echebarria \emph{et al.} \cite{Echebarria04} have shown that mesh
converged results can be obtained with $\epsilon$ as large as 50.
Setting $\epsilon = 50$, gives us an under-determined system of
three equations with the five unknowns $D$, $V_p$, $l_T$, $\tau_0$
and $W_0$. Making arbitrary choices for two of these parameters by
setting $W_0 = \tau_0 = 1$, we obtain $D = 27.7$, $V_p = 0.2825$,
and $l_T = 300$. A large value of $l_T$ implies $L_x$ needs to be
very large at steady state, even in a moving reference frame, to
contain the diffusion field. To avoid this, if we choose a more
tractable value for $l_T$ (say 100), and fix $\tau_0 = 1$, we now
get $W_0 = 0.3333$, $D = 3.077$ and $V_p = 0.094$. Thus, the
smallest element in our mesh needs to be at the very least $\Delta
x = 0.3333$. Stability considerations now place a severe
restriction on the size of the time step ($\Delta t$) needed for
solving the phase-field equations by the Forward-Euler method, as
$\Delta t \sim \Delta x^2$ . Since it is clearly impossible to
choose both $l_T$ and $W_0$ independently, calculations involving
real materials are typically more expensive.

We choose instead a more computationally favorable set of
dimensionless parameters $M$ and $S$ for our study. To achieve our
primary objective, which is to study the effect of $\delta$ on
interface morphology, we anticipate, and in the following
paragraphs demonstrate, that this hypothetical treatment will not
obscure any physics. The parameters used in our study are: $\tau_0
= W_0 = 1$, $D = 20$, $k=0.8$, $\epsilon_4 = 0.05$, and $l_T =
|m|(1-k)C_{\infty}/kG = 50$; where $m$ is the liquidus slope, $G$
is the imposed thermal gradient, and $C_{\infty}$ is the far field
solute concentration. The condition for negligible interface
kinetics gives $d_0 = 0.0277$ and therefore $\epsilon = 36.1$,
which is sufficiently small to ensure convergence. The size
of the smallest element in our adaptive mesh is $\Delta x = 1$,
when $\delta \ge 8$, and $\Delta x = 0.5$, when $\delta = 4$,
while $\Delta t = 0.005$ is the size of the time-step. For these
parameter choices, the dimensionless control parameters work out
to be $M = 5.54 \times 10^{-4}$ and $S \sim 1.385 \times 10^{-3}$.

\subsection{Interface morphology and comparison with experiments}

In order to test the model, we initially performed a set of runs
to verify the Mullins and Sekerka stability limit of a planar
interface \cite{mullins}, perturbed by small sinusoidal
perturbations. We found that the model captures the stability
spectrum correctly. Having convinced ourselves that this
fundamental requirement was met, we proceeded with our study.
Comparisons with the experimental data in this section offer a
better validation of the model.

\begin{figure}[htbp]
\begin{center}
\subfigure[$\delta = 64$, $V_p = 1.5$, three-dimensional cells
\label{ds_pix_a}]
{\includegraphics[height=4.0in,angle=90]{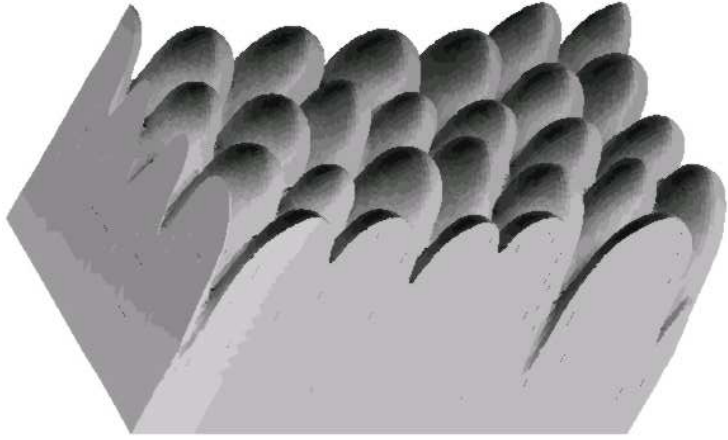}}
\subfigure[$\delta = 16$, $V_p = 1.5$, three-dimensional cells
\label{ds_pix_b}]
{\includegraphics[height=4.0in,angle=90]{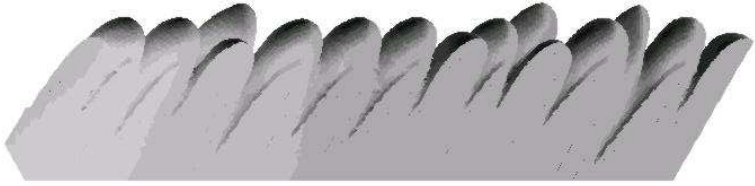}}
\subfigure[$\delta = 4$, $V_p = 0.8$, two-dimensional cells
\label{ds_pix_c}]
{\includegraphics[height=4.0in,angle=90]{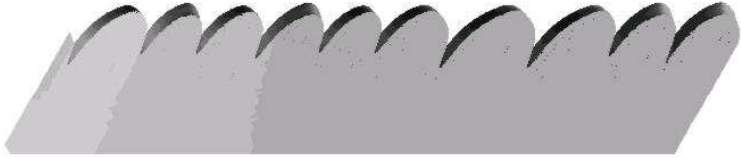}}
\caption{\label{ds_pix} Interface morphology in a directionally
solidified alloy. At steady state, stable arrays of
three-dimensional cells appear. Note that at $\delta = 4$, the
array comprises of two-dimensional cells.}
\end{center}
\end{figure}

Fig. \ref{ds_pix} shows the computed interface morphology at
different values of $\delta$. For small values of pulling speed
($V_p \le 2$), the steady state consists of a stationary array of
cells as in Fig. \ref{ds_pix}. However, unlike the cells observed
in experiments, that are usually characterized by blunt tips,
these appear to have sharper and better defined tips, giving the
impression of dendrites. It is conceivable that ignoring thermal
noise in our calculations is responsible for the absence of
side-branches on these structures, that are typical of dendrites.
At large values of $\delta$, we find that the tip radii of these
cells, measured on the two principal planes, are almost identical.
However, as $\delta$ decreases, the in plane radii diverge from
one another. In particular, the radius in the \emph{x-z} plane
becomes significantly smaller, and cross sections of the cells
look elliptic. At $\delta = 4$, for small pulling speeds ($V_p \le
1$), we get a two-dimensional interface (Fig. \ref{ds_pix_c}). The
inter-cellular spacing also increases as $\delta$ is decreased.

As pulling velocity is increased, the morphology becomes finer,
with sharper and more tightly packed cells. This behavior is
consistent with that seen of both cellular and dendritic arrays in
directional solidification experiments
\cite{Somboonsuk84,kirkaldy+liu,liu+trivedi}, where the primary
spacing decreases with $V_p$.


For the set of phase-field parameters we have chosen, if $V_p \ge
2.5$, the interface does not reach steady state in a reasonable
amount of CPU time, due to repeated tip-splitting of the cells.
Splitting is initiated by oscillations that appear at the tip and
propagate downward along the trunk of the cell. Cell spacing and
shapes change very rapidly in this regime. We did not continue
these runs any further to check if steady state is reached
eventually. Instead, we set $V_p = 2.0$ as an upper bound on the
pulling velocity, below which a stationary state was always the
outcome. Fortunately, this still left us with sufficient sample
space to conduct our study and make effective comparisons.

\begin{figure}[htbp]
\begin{center}
{\includegraphics[height=4.0in,angle=0]{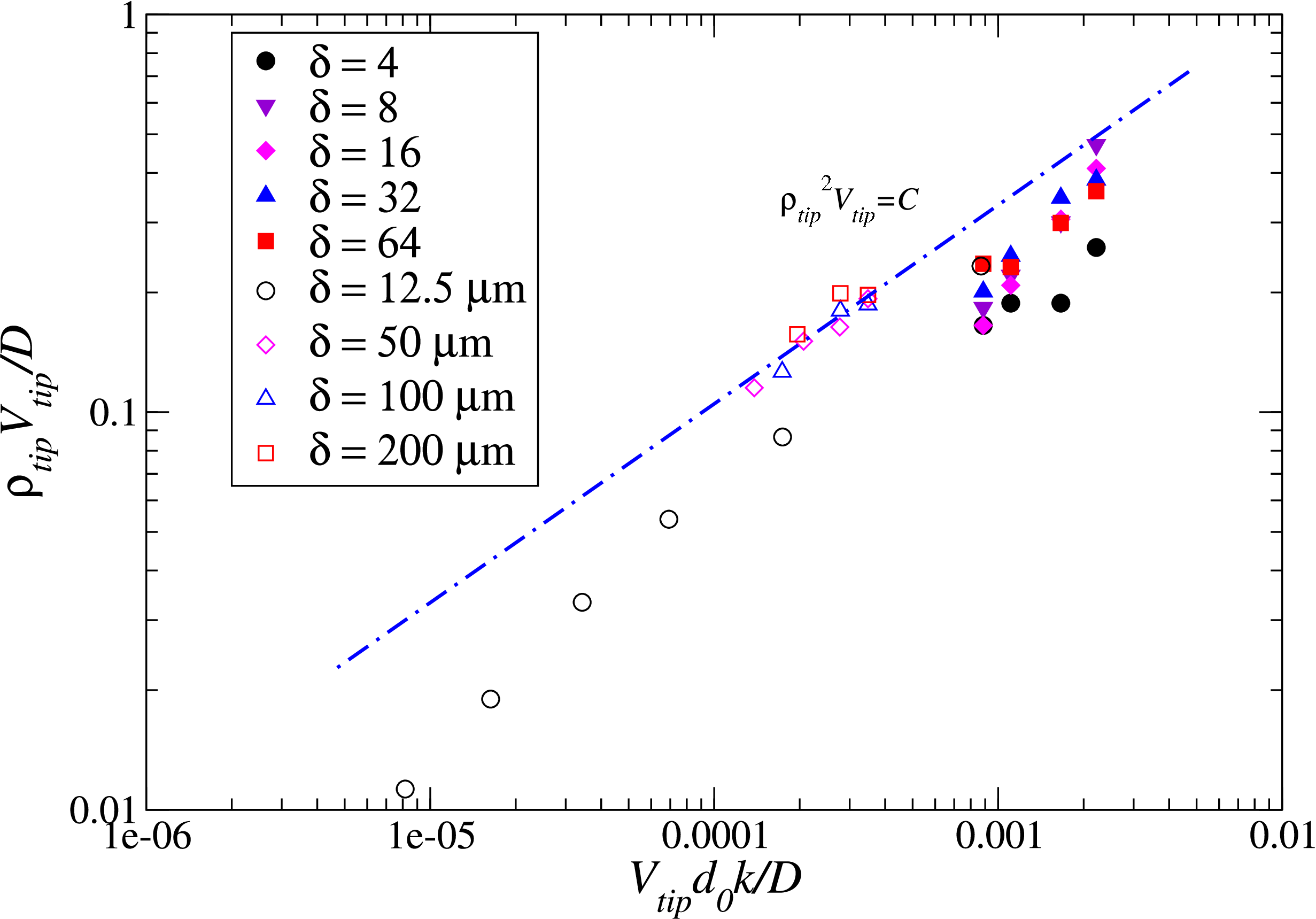}}
\caption{\label{petip_comp} Comparison of binary alloy simulations
with the phase-field model, and experimental data of Liu \emph{et
al.} \cite{liu+trivedi}. Solid symbols correspond to the
phase-field model and the open symbols are experimental data. The
line denotes a relationship of the form $\rho_{tip}^2V_{tip} = C$
between tip radius and velocity, where $C$ can be expressed in
terms of process parameters.}
\end{center}
\end{figure}

To enable plotting of our results on the same graph with the
experimental data of Liu \emph{et al.}, which corresponds to a SCN
- 0.7 \% wt. Salol system (properties in Table 3), we
non-dimensionalized the axes as follows. The abscissa is the
pulling speed (i.e. the tip velocity at steady state) $V_{tip}$,
scaled by a characteristic velocity $D/d_0k$, while the ordinate
is the tip radius $\rho_{tip}$, scaled by the diffusion length
$D/V_{tip}$. One may appreciate the fact that the abscissa is in
fact the dimensionless parameter that we had earlier denoted by
$S$, multiplied by $k$. Fig. \ref{petip_comp} shows a comparison
of the data. The open symbols correspond to the experimental data
of Liu \emph{et al.}, while the solid symbols correspond to our
calculations. A comparison between our data and theirs holds up
surprisingly well. Of special significance are the following two
observations: 1) Although we conducted our simulations at values
of $S$ and $M$ that were each about an order of magnitude off
theirs, the two sets of data correlate very well, i.e. appear to
collapse on parallel curves that are not significantly different
by way of intercept. This tells us that our choices of parameters
for scaling the axes are appropriate. 2) Since their experimental
data correspond to dendritic arrays, the cell-like structures we
have computed are likely branchless dendrites.


In their experiments, Liu \emph{et al.} note that tip radius data
for dendritic arrays agree quite nicely with a relationship of the
form $\rho_{tip}^2V_{tip} = C$, where $C$ is a constant dependent
on $d_0$, $k$ and $D$, as postulated by theoretical models of
constrained growth \cite{Trivedi80,Kurz+Fisher}. However, when
$\delta$ is of the order of inter-dendritic spacing $\lambda_1$,
this agreement deteriorates. This is evident in Fig.
\ref{petip_comp}, where the $\rho_{tip}^2V_{tip} = C$ line is
shown as a guide to the eye. The open circles, which are data for
$\delta = 12.5\;\mu$m deviate in both slope and intercept from
this line, which passes through the rest of their data, indicating
a breakdown in the relationship. We observe similar trends in our
data, viz., the line $\rho_{tip}^2V_{tip} = C$ fits our data at
$\delta = 32\;\mbox{and}\; 64$ reasonably well, but as $\delta$
decreases from 16 to 4, this agreement deteriorates.

As in the numerical experiments with the pure material, we find
that decreasing $\delta$ has a pronounced effect on interface
morphology. Dendritic arrays seen in experiments have a certain
structure/periodicity to them, that arises from underlying
crystalline symmetries. For example, in our simulations we observe
that the cells constitute a hexagonal array. When $\delta$ is
large, away from the boundaries the diffusion field surrounding
each cell tip obeys this symmetry, and the optimal $\lambda_1$ is
selected. As $\delta$ decreases however, the diffusion field
becomes increasingly asymmetric due to interaction with the
boundaries at $x=0$ and $x=\delta$. In particular, solute
rejection decreases in the vertical plane \emph{x-z}, while
increasing in the horizontal plane \emph{x-y}. Increased solute
accumulation between cells in \emph{x-y}, contributes to an
increase in $\lambda_1$. However, since $V_{tip}$ is fixed by the
pulling speed, and a certain rate of solute rejection needs to be
maintained, the tips tend to grow sharper as $\lambda_1$
increases. It is precisely this effect that causes the operating
state to deviate from theoretical predictions.

To check for what value of $\delta$ our results deviate from
theory, let us assume the steadily growing array at $\delta = 64$,
$V_p = 1.5$ (Fig. \ref{ds_pix_a}) to be one in which cells close
to the plane $z = 32$ are ``free'' from boundary effects. We
estimate the cell spacing in that plane to be $\lambda_1^f \approx
18$, where the superscript `f' denotes ``free''. When $\delta =
16$ and $V_p=1.5$ (Fig. \ref{ds_pix_b}), we notice that our data
points start departing from the theoretical prediction, viz. we
see a relationship of the type $\rho_{tip}^aV_{tip} = C$, where $a
> 2$. This suggests that agreement with theory deteriorates as $\delta
\sim \lambda_1^f$, which is what Liu \emph{et al.} concluded from
their experiments. A more precise form of the above criterion can
be obtained my making a careful study of $\lambda_1$ as a function
of $\delta$, for different $V_p$, and obtaining a criterion based
on a least squares fit to the deviation points (as we did with the
pure material). Increasing pulling velocity suppresses
tip-boundary interaction by reducing the thickness of the
diffusion boundary layer $D/V_{tip}$, similar to the effect of
melt convection in pure materials, and should induce a leftward
shift in the curves.

\section{Concluding remarks \label{s5}}

We have investigated the role of confinement on solidification in
both pure materials and binary alloys. Our simulations show that,
for equi-axed growth in a pure material, the dendrite's operating
state is affected when the container dimension $\delta$,
approaches the scale of the diffusion field $D/V_{tip}$ near the
tip. For directionally solidified binary alloys, confinement
effects become important when $\delta$ is of the order of the
primary dendrite spacing $\lambda_1$. Where applicable, one needs
to consider the influence of these interactions when comparing
experimental data with theoretical models that do not account for
confinement effects.

It is notable  that we were able to make
meaningful comparisons with real experimental results using the
phase-field model for the alloy. In particular, the agreement
obtained in the trends shown by the dendrite tips for different
$\delta$ and $V_p$ is very encouraging, and is a testament to the
power of phase-field modeling. Some ambiguity remains in
classifying the computed microstructure as cells or dendrites.
Since our results correlated well with dendrite data, we would
like to think of them as dendrites. Perhaps, incorporating random
fluctuations in the phase-field model will resolve this issue. We
also found tip-splitting inducing oscillations above certain
values of the pulling speed. We are unsure as to whether this
instability has a physical meaning. In experiments, it is seen
that increasing $V_p$ causes a decrease in $\lambda_1$ and
$\rho_{tip}$ for a steadily growing interface, and our simulations
capture this effect, viz. as $V_p$ increases the cells split in a
manner that produces a more stable configuration with a smaller
$\lambda_1$ and $\rho_{tip}$. However, when $V_p \ge 2.5$, it
appears that an optimal configuration is not possible in our
system, given the constraints on its size and boundary conditions.
We speculate that larger domains should allow for more stable cell
configurations at higher pulling speeds. This issue too needs
further investigation.

We observed a change in dimensionality of the liquid-solid
interface for certain values of $\delta$ and $V_{tip}$, when zero
flux boundary conditions were imposed on the phase-field variable.
There is some experimental evidence of this phenomenon in the
literature. Liu and Kirkaldy \cite{kirkaldy+liu} reported a 2-D to
3-D transition in their experiments on a SCN-Salol mixture. In
their directional  solidification experiments in a cell of fixed
height ($\delta = 28 \mu$m),  they found this transition to occur
at a driving velocity of 10.8 $\mu$m/sec.  At a lower driving
velocity of 7.6 $\mu$m/sec, the dendrites looked two dimensional.
In our analysis of directional solidification, we found at $\delta
= 4$, the cells underwent a 2-D to 3-D transition as the pulling
velocity was changed from 1 to 1.5. The significance of this
result is that through an appropriate selection of $\delta$ and
$V_p$ in experiments, it should be possible to obtain almost two
dimensional dendritic arrays in materials that favor wetting. Such
experiments will permit more favorable comparisons with 2-D
dendrite growth theories, since finite boundary effects along the
\emph{z} axis cease to impact the growth.

\begin{acknowledgements}
The authors gratefully acknowledge support for this work from NASA
under Grant NAG 8-1657, and from the National Science Foundation
under Grant DMR 01-21695. We thank Jun-Ho Jeong and Navot Israeli
for their significant contributions in the development and
implementation of the adaptive grid algorithm which made this
study possible, and the Computational Science and Engineering
program at UIUC for availing us their computing facilities.
\end{acknowledgements}

\bibliographystyle{apsrev}

\end{document}